\documentstyle[aps]{revtex} 
\author{J. Rotyis\cite{*}{\dag} and G. Vattay{\ddag}\\ 
{\dag} FOCUS Investment Rating Company, H-1053 Budapest, Sz\'ep u. 2 X/1010, Hungary\\ 
{\ddag}E\"otv\"os University, Physics of Complex Systems Group, H-1088 Budapest,  
M\'uzeum krt. 6-8, Hungary \\ vattay@ludens.elte.hu} 
\title{Statistical Analysis of the Stock Index of the Budapest Stock Exchange} 
\begin{document} 

\input psfig\wideabs{ 
\maketitle 
\begin{abstract} 
Scaling properties of the BUX index are 
similar to those observed in other parts of the world. The 
main difference is that the traditional quantities 
like volatility, growth and autocorrelation of returns follows 
more closely the assumptions of the traditional stock market 
theory developed by Bachelier and by Black and Scholes. 
\end{abstract} }
 
\section{Historic notes} 
 
The Budapest Stock Exchange\cite{Szaz} (BSE) opened in 1864 and had its best  
period around the turn of the century, when it become the fourth largest 
stock exchange in Europe. It was closed temporarily in 1945 due to the 
siege of Budapest and in 1919 for few months and 
in 1948 permanently as a consequence of  
communist takeovers in Hungary. Then in 1982 a new legislation permitted 
again the issue of bonds. In 1988 a semi-official trading of bonds and 
securities started again and it become possible to establish share companies.  
The Exchange has been reestablished on 19 of June 1990 
and had its first trading day on 21 of June.  
 
The base day of the index of the BSE, the `BUX', is 2 of January 1991 
when it was set to 1000 points. It is calculated through a weighted average of 
the prices of main stocks and represents the total capitalization of the 
market. The BUX index reached its minimum in May 1993 at 718 points, 
caused by the escalation of the war in ex-Yugoslavia. 
The index, after some ups and downs reached the value 1500 by 
the end of 1995. Since then the index went up to about 8000 in 
August 1997 steadily, producing the highest (inflation adjusted) 
stock exchange growth in the world in 1996. This success is expected 
to be repeated in 1997. 
 
In this paper we analize the index and some of the main 
stocks for the period January-June 1997 a period of steady growth. 
 
\section{The data}   
 
In the present analysis we use the data provided by the BSE. 
Securities trading on 
the BSE is entirely electronic. Dealers submit their ask and bid 
prices, the amounts and the computer system matches the orders between 
partners. The computer gives priority to higher bids and lower asks
and under equal conditions time of orders also counts.
The time, the amount 
and the price is then recorded. We have such records for the five most 
traded stocks of the BSE for the entire period 6 of January - 30 of 
June 1997 for each trading day.  
 
Before 1 of April 1997 the index was calculated at the end of  
each trading day based on the market closing prices. Since then the index is 
calculated continuously. On each treading day the index calculation begins 
15 minutes after the opening of the market. Then it is 
calculated in each 5 seconds. Very often no deal is made within 5 seconds  
and the index remains unchanged. We also have the BUX index record for 
the period 1 of April - 30 of June 1997. In Fig. 1 we show the evolution 
of the real time BUX index for this period.  
 
\begin{figure} 
\centerline{\strut\psfig{figure=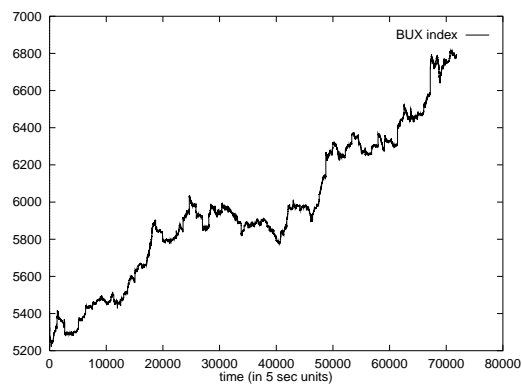,width=50mm}} 
\caption{Real time BUX index for the period 1 of April - 30 of June 
1997. We joined the daily records and produced a single data sequence. 
The time is measured in 5 seconds units.} 
\label{fig1} 
\end{figure}

\section{Scaling and distribution of returns} 
 
Following Refs.\cite{breymann,stanley} we study the scaling properties of 
returns of the BUX index. The return over $n$ time steps is defined as 
\begin{equation} 
Z_n(t)=X(t+n\Delta t)-X(t), 
\end{equation} 
where $X(t)$ is the BUX index and $\Delta t=5 sec$ is the sampling time. 
We have found that the moments of the distribution of $Z_n$ show 
scaling behavior as a function of $n$ 
\begin{equation} 
\langle |Z_n(t)|^q \rangle_t \sim n^{\xi_q}, 
\label{scala}
\end{equation} 
where $\xi_q$ is the self-affinity exponent. 
\begin{figure} 
\centerline{\strut\psfig{figure=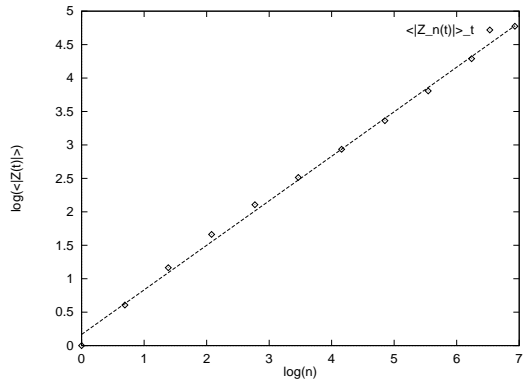,width=50mm}} 
\caption{Scaling of the absolute value of the return. {\em Natural 
logarithms} of the quantities are plotted. The scaling $\sim n^{0.666}$ 
is observed for almost three decades.} 
\label{fig2} 
\end{figure} 
In Fig. 2 we show the $q=1$ moment for 
$n=1,...,1024$ . We can observe scaling for three decades 
(on decimal scale). We have found that this remarkable agreement with the 
scaling assumption (\ref{scala}) holds for $q$ values larger than $0.2$ . 
Traditional stock market theory\cite{bachelier} (TSMT) predicts 
Brownian motion for 
the stock prices which leads to a uniform exponent $\xi_q=q/2$. 
In Fig. 3 we show the measured exponents $\xi_q$ as a function of $q$. 
\begin{figure} 
\centerline{\strut\psfig{figure=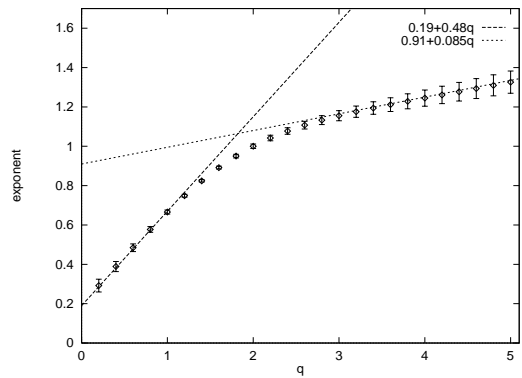,width=50mm}} 
\caption{Measured values of $\xi_q$ for the BUX index. The errorbars  
represent the error of the fit of the power law.} 
\label{fig3} 
\end{figure} 
The function $\xi_q$ is strongly non-linear, 
differs strongly from the behavior of the S\&P 500 index 
in Ref.\cite{stanley} and DM/US\$ exchange rate in Ref.\cite{breymann}. 
 
We can see that the exponent for $q=1$ is close to $2/3$ deviating 
significantly from TSMT, while for $q=2$ the exponent is about $1$ 
predicted by TSMT. This is very puzzling, since traditionally the 
volatility is measured via the relation 
\begin{equation} 
<|X(t+n\Delta t)-X(t)|^2>\sim\sigma^2 n, 
\end{equation} 
which seems to hold for the BUX index, despite the strongly 
non-Brownian behavior of other cummulants.  
 
It is even more  
surprising that the growth rate seems to follow TSMT. 
In Fig. 4 we show the average behavior of 
\begin{figure} 
\centerline{\strut\psfig{figure=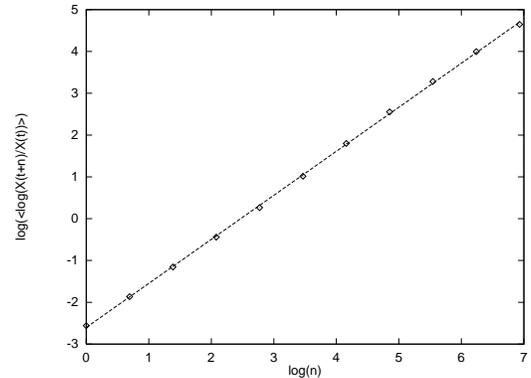,width=50mm}} 
\caption{Average value of $<\log(X(t+n\Delta t)/X(t))>$ for 
$n=1,2,...,1024$ and  $\Delta t= 5 sec$. The best fit line 
represents $X(t+n\Delta t)=e^{0.074n}X(t)$.} 
\label{fig4} 
\end{figure} 
$\log X(t+n\Delta t)-\log X(t)$. We have found that 
\begin{equation} 
\langle\log(X(t+n\Delta t)/X(t))\rangle \approx \mu n , 
\end{equation} 
where $\mu\approx 0.074=e^{-2.60}$ so that the expected behavior of 
the index is 
\begin{equation} 
X(t+n\Delta t)\approx X(t)e^{\mu n}. 
\end{equation}  
 
Next, we studied the distribution $P_n(Z)$ of the returns $Z_n(t)$ 
shown in Fig. 5.  
\begin{figure} 
\centerline{\strut\psfig{figure=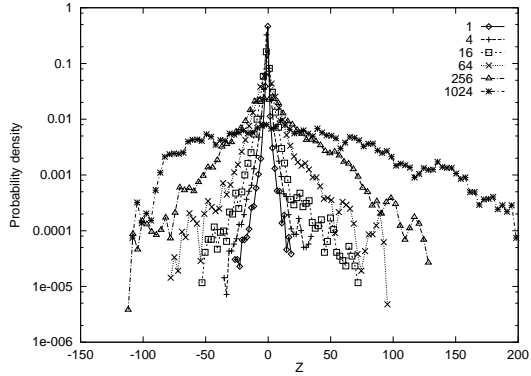,width=50mm}} 
\caption{The probability density $P_n(Z)$ on log-linear plot
for $n=1,4,16,256,1024$.} 
\label{fig5} 
\end{figure} 
These curves are similar to those observed  
for the S\&P index in Ref.\cite{stanley}. The center of the 
distribution shows similar 
scaling $P_n(0)\sim n^{-1/\alpha}$, where the exponent is  
$1/\alpha\approx 0.666$, $\alpha\approx 1.50$ in accordance with  
the value $\xi_1\approx 0.66$ and close to the S\&P value 
$\alpha=1.40$ of Ref.\cite{stanley} and coincides with the 
exponent found by Bouchaud et al. in Ref.\cite{bou}.   
The distribution, however, cannot be fitted with a L\'evy type 
distribution since the rescaled curves 
$P_n(Z/n^{0.66})n^{0.66}$ do not collapse on a single curve. 
This fact is in accordance with the strong non-linearity 
observed in $\xi_q$. We can say qualitatively that the  
distributions are similar to L\'evy distributions with parameter 
$\alpha=1.5$ up to times $\approx 300 sec$ and Gaussian (log-normal) 
distribution sets in around $\approx 5000 sec$. 
 
Subsequently we studied also the scaling behavior of the main stocks 
on BSE which lead to similar exponents and conclusions. 
 
\section{Correlations and Fourier analysis}  
 
Temporal correlations or their absence plays an important role 
in the predictability of the market. Here we study the behavior  
of the correlation functions 
\begin{equation} 
C_n(m)=\langle Z_n(t+m\Delta t)Z_n(t)\rangle_t . 
\end{equation} 
According to TSMT short time increments of stock prices are 
uncorrelated so the normalized autocorrelation function of 
the single increments is 
\begin{equation} 
C_1(m)/C_1(0)=\delta_{0,m}. 
\end{equation} 
We tested this hypothesis for the BUX index and have found 
perfect agreement. The absence of short time correlations can 
be the result of the random match making process between dealers 
mentioned in the previous section. Then  
the increments $X(t+n\Delta t)-X(t)$ can be considered as a  
sum of $n-1$ uncorrelated steps. The shifted series $X(t+(m+n)\Delta t) 
-X(t+m\Delta t)$ is composed of $n-1$ uncorrelated increments, however 
$n-1-m$ of these increments coincide with those in $X(t+n\Delta 
t)-X(t)$. As a consequence, the TSMT prediction for $C_n(m)$ is 
\begin{eqnarray} 
C_n(m)/C_n(0)&=&1-m/(n-1) \;\;\;\mbox{for $m<n$},  \nonumber \\ 
&=& 0 \;\;\;\mbox{otherwise.} \label{cori} 
\end{eqnarray}  
In Fig. 6 we show the absolute value of the autocorrelation functions 
\begin{figure} 
\centerline{\strut\psfig{figure=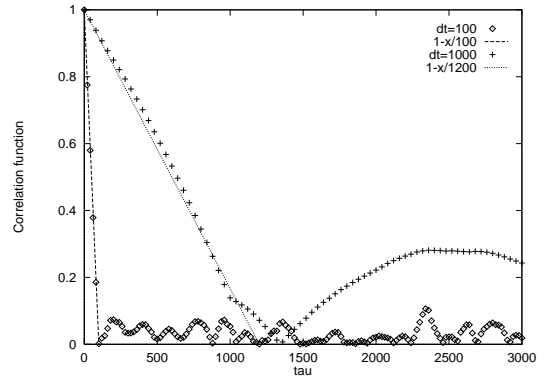,width=50mm}} 
\caption{Autocorrelation functions $|C_{100}(\tau)/C_{100}(0)|$ and 
$|C_{1000}(\tau)/C_{1000}(0)|$. Notice, that fluctuations  
of the $n=1000$ correlation function are about 
$(10)^{1/2}$ times larger than those for $n=100$ because of the 
higher redundancy of the series $Z_{1000}(t)$. The fluctuating 
parts are insignificant in both cases.} 
\label{fig6} 
\end{figure} 
for the BUX index for $n=100$ and $n=1000$. We have found that  
correlation functions up to $n\approx 100$ behave in accordance 
with (\ref{cori}) while above that correlations are higher than 
those predicted by TSMT. For example, the $n=1000$ ($5000 sec$)  
data can be 
well fitted with $C_{1000}(m)/C_{1000}(0)=1-n/1200$. This increased level 
of correlations can be observed by looking at the power spectrum 
\begin{equation} 
S_n(f)=\left|\int dt e^{i2\pi f t}Z_n(t)\right|. 
\end{equation} 
For $n<100$ the power spectrum is constant indicating that the 
series $Z_n(t)$ is white noise. For $n\approx 1000$ the  
power spectrum shows scaling behavior (see Fig. 7) and 
\begin{figure} 
\centerline{\strut\psfig{figure=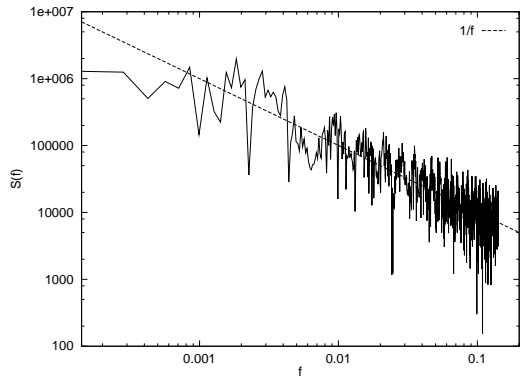,width=50mm}} 
\caption{Power spectrum $S_{1000}(\omega)$ of $Z_{1000}(t)$. The solid 
line indicates $1/f$.} 
\label{fig7} 
\end{figure} 
it is very close to  
\begin{equation} 
S(f)\sim 1/f, 
\end{equation} 
indicating the presence of noise related to one-dimensional
Brownian motion.  
 
\section{Conclusions} 
 
As we demonstrated, the scaling properties of the BUX index are 
similar to those observed in other parts of the world. The 
main difference is that the traditional quantities 
like volatility, growth and autocorrelation of returns follows 
more closely the assumptions of the traditional stock market 
theory developed by Bachelier and by Black and Scholes. 
Bouchaud argued\cite{bou} that short time correlations are in 
accordance with the efficient market hypothesis provided  
transaction costs makes it impossible to use them for arbitrage. 
Since there was no transaction fee on the BSE during the period of our 
study\cite{nn} this might be responsible for the complete absence of  
short time correlations observed elsewhere. The reason for long time 
(5000 sec) correlations is less clear and a more 
extended study is needed to find its origin. 
 
The authors thank Ildik\'o Fecser, Tam\'as Laskai and Zsolt Kendi,
the team of the BSE, for their support of this study. G. V. thanks
I. Csabai, B. Janecsko, I. M. J\'anosi and I. Kondor the
discussions and the financial support  of the Hungarian Science 
Foundation OTKA  (F019266/F17166/T17493) and the Hungarian Ministry of 
Culture and Education.

\end{document}